\documentclass[manuscript,graphicx]{aastex}

\begin{document}

\voffset = 0.0truecm

\newcommand{\gray}{$\gamma$-ray\  } \newcommand{\grays}{$\gamma$-rays\
} \newcommand{\etal}{{\it  et al.\ }}  \newcommand{\lya}{Ly$\alpha$\ }
\newcommand{\epr}{e-print  astro-ph/  }  %\newcommand{\e} {\epsilon  }
\newcommand{\mic}{$\mu$m\ } %\slugcomment{Astrophysical
%Journal Letters 652, L9 (2006)}
\title{Derivation  of a Relation  for the  Steepening of  TeV Selected
Blazar $\gamma$-ray Spectra with Energy and Redshift}

\author{Floyd   William  Stecker}  \affil{NASA/Goddard   Space  Flight
Center}           \authoraddr{Greenbelt,           MD           20771}
\email{Floyd.W.Stecker@nasa.gov}      \author{Sean     T.      Scully}
\affil{Department    of    Physics,    James    Madison    University}
\authoraddr{Harrisonburg, VA 22807} \email{scullyst@jmu.edu}

\begin{abstract}

We derive a relation for the steepening of blazar $\gamma$-ray spectra
between  the multi-GeV  {\it Fermi}  energy range  and the  TeV energy
range observed by atmospheric  \v{C}erenkov telescopes.  The change in
spectral  index   is  produced  by  two  effects:   (1)  an  intrinsic
steepening,  independent  of  redshift,  owing to  the  properties  of
emission and  absorption in the  source, and (2)  a redshift-dependent
steepening produced  by intergalactic pair  production interactions of
blazar $\gamma$-rays  with low  energy photons of  the ``intergalactic
background light''  (IBL). Given this relation, with  good enough data
on the mean \gray SED of  TeV Selected BL Lacs, the redshift evolution
of the IBL  can, in principle, be determined  independently of stellar
evolution models.   We apply our relation  to the results  of new {\it
Fermi} observations of TeV selected blazars.
\end{abstract}


\section{Introduction}

Stecker \&  Scully (2006) (SS06) derived a  simple analytic expression
for the change in spectral index of a TeV \gray source in the redshift
range  between 0.05  and  0.4.  They  showed  that the  change in  the
spectral  index caused  by  intergalactic absorption  is  given by  an
approximately  linear  relation   in  redshift,  {\it  i.e.},  $\Delta
\Gamma_{a} \simeq C + Dz$.

The purpose of this letter  is to generalze this relation by including
the effect of  intrinsic steepening in the source  spectra between the
mutli-GeV  energy range  observed by  {\it  Fermi} and  the TeV  range
observed by atmospheric  \v{C}herenkov telescopes.  Our general result
is  roughly independent  of the  specific model  of  the intergalactic
background light (IBL)  used, because it only depends  on the shape of
the average galaxy  spectral energy distribution (SED) on  the near IR
side of the  starlight peak that determines the  absorption in the TeV
energy range.

We compare our  relation for the specific baseline  and fast evolution
models of Stecker, Malkan \& Scully (2006) (SMS06) with the results of
recent {\it Fermi} observations of 13 TeV selected BL Lac AGN. We also
show  how it  can  be  used to  independently  determine the  redshift
evolution of the IBL.

\section{Steepening by Absorption}

In order to  determine the effect of intergalactic  absorption, we use
the results  of SS06 demonstrating  that $\tau(E_{\gamma}, z)$  can be
fitted to an approximately logarithmic function in $E_{\gamma}$ in the
energy range $0.2~ \rm  TeV < E_{\gamma} < 2 ~\rm TeV  $ and one which
is linear on $z$  over the range $0.05 < z <  0.4$. It is important to
note  that  our  linear  fit  to  the  $z$  dependence  is  both  {\it
qualitatively}  and  {\it quantitatively}  different  from the  linear
dependence on redshift which would  be obtained for small redshifts $z
<< 1$  and which simply comes from  the fact that for  small $z$ where
luminosity evolution  is unimportant and where $\tau  \propto d$, with
the distance $d  \simeq cz/H_{0} \propto z$. Our  quantitative fit for
the higher redshift range $0.05 < z < 0.4$ comes from the more complex
calculations based on  the models of SMS. For  this reason, the linear
fits in SS06 are not simply proportional to redshift.

SS06 found that $\tau(E_{\gamma},z)$ is well approximated by

\begin{equation}
\tau(E_{\gamma},z) = (A + Bz) + (C + Dz)\ln [E_{\gamma}/(\rm 1~ TeV)],
\end{equation}

\noindent where  $A, B,  C ~$and $D$  are constants.   This expression
holds over  the energy  and redshift ranges  given above.   The energy
range  of  validity is  the  energy  range  to which  the  atmospheric
\v{C}erenkov telescopes are sensitive.

Following SS06,  we assume  an intrinsic source  spectrum that  can be
approximated  by a  power law  of the  form $\Phi_{s,TeV}(E_{\gamma})~
\simeq~  kE_{\gamma}^{-\Gamma_{s,TeV}}$ over  a limited  energy range.
Then  the  spectrum that  will  be  observed  at the  Earth  following
intergalactic absorption will be of the power-law form

\begin{equation}
\Phi_{o,  TeV}(E_{\gamma})~  = ~  ke^{-(A+Bz)}E_{\gamma}^{-(\Gamma_{s,
TeV }+C+Dz)}.
\end{equation}

This can  be compared with  the empirically observed TeV  spectra that
are usually presented in the literature to be good fits to power-laws.
The observed spectral index, $\Gamma_{o,TeV}$, will then be given by

\begin{equation}
\Gamma_{o, TeV} = \Gamma_{s, TeV} + \Delta \Gamma_{a, TeV} (z)
\end{equation}

\noindent  where  the  intrinsic  spectral  index  of  the  source  is
steepened by intergalactic absorption by an amount $ \Delta \Gamma_{a}
(z)~ =~ C + Dz$.

On the other  hand, in the multi-GeV range over  which the {\it Fermi}
Large  Area Telescope  (LAT) is  sensitive, we  expect  essentially no
steepening from  absorption over the redshift range  of validity, {\it
i.e.}, $\Delta \Gamma_{a, GeV} (z) \simeq 0$.

The parameters  $ C$, and $D$  obtained by fitting  the optical depths
derived for the  fast evolution (FE) and baseline  (B) models of SMS06
are given in Table 1.

\section{Intrinsic Steepening} 

The importance of synchrotron  self Compton emission in producing high
energy \grays in astrophysical sources was pointed out by Rees (1967).
Particular applications  to the Crab and other  sources were discussed
by  Gould (1965)  and Rieke  \& Weekes  (1969).  Today,  given present
observational  studies of  TeV-selected BL  Lac AGN,  it  is generally
accepted that  the synchrotron self  Compton mechanism is  the primary
emission mechanism  for producing  \grays in the  TeV energy  range in
TeV-selected BL Lac AGN.

The \gray spectrum from Compton  interactions is a smoothy varying one
(Blumenthal \& Gould  1970).  Good empirical fits the  the SEDs of the
low energy peaks in blazars were obtained by using parabolic functions
on  log-log plots  (Landau, et  al. 1986,  Sambruna, Maraschi  \& Urry
1996).   In  fact,  these  considerations  led to  the  prediction  of
candidate  BL Lac  objects  to be  found  by atmospheric  \v{C}erenkov
telescopes in the TeV energy range (Stecker, de Jager \& Salamon 1996;
Costamante \& Ghisellini 2002).

We assume such an approximation here, {\it i.e.},

\begin{equation}
E^2 {{dN_{\gamma}}\over{dE}} = f(x),
\end{equation}

\noindent where  $x = \log E$  with the simplification that  $E$ is in
dimensionless fiducial units ({\it i.e.},$~E/E_f$), and

\begin{equation}
f(x) = {-{(x-x_0)^2}\over{2A}} + B.
\end{equation}

Here,  $A  \equiv \log  <W>$  is a  constant  parameter  based on  the
ensemble averaged width  of the parabola in $\log E$  space based on a
set of  SEDs of  TeV selected  BL Lacs and  $x_{0} \equiv  \log E_{0}$
where $E_{0}$ is  the energy at which the Compton  peak of the average
BL  Lac SED  is a  maximum. We  note that  for some  BL  Lacs electron
scattering  in the  Klein-Nishina  regime will  distort our  parabolic
assumption on  the high  energy end of  their SEDs,  possibly reducing
$\log <W>$ by a small amount (Tavecchio, Maraschi \& Ghisellini 1998).
However, this does not significantly affect our formalism.

We find that the change in  the slope of the intrinsic source spectrum
occurring  between  some average  observed  energy  in  the GeV  range
$<E_{\rm  GeV}>$ and  some average  observed energy  in the  TeV range
$<E_{\rm TeV}>$ is approximately given by a constant,

\begin{equation}
{{\log <E_{TeV}> - ~\log <E_{GeV}>}\over{A}} \equiv K
\label{K}
\end{equation}

Thus, the ensemble averaged intrinsic  BL Lac source spectrum in going
from the GeV range to the  TeV range is independent of redshift within
the errors  in the observational  data. We will  therefore approximate
the average intrinsic steepening over  an observed set of TeV selected
blazars by

\begin{equation}
\Gamma_{s, TeV} - \Gamma_{s, GeV} = \Delta \Gamma_{s} = K.
\end{equation}

\section{The Relation for the Total Steepening between GeV and TeV Energies}

The total steepening expected betwen  the GeV and TeV energy ranges is
then given by the relation

\begin{equation}
\Delta \Gamma = \Delta \Gamma_{s} + \Delta \Gamma_{a} = (C + K) + Dz
\label{total}
\end{equation}

The parameters  $ C$, and $D$  obtained by fitting  the optical depths
derived for the  fast evolution (FE) and baseline  (B) models of SMS06
are given in Table 1. The  parameter $K$ is then derived by performing
a $\chi ^2$  fit to the observed steepening data  obtained by the {\it
Fermi}  collaboration (Abdo,  et  al. 2009)  in  order to  find $(C  +
K)$. The  parameter $K$  is also  given in Table  1 for  the FE  and B
models. The data on the TeV  selected BL Lacs from Abdo, et al. (2009)
are shown in figures 1 and 2, along with the best-fit linear relations
for  the models indicated.   It can  be seen  that, given  the present
limited data set  and large error bars, one  cannot uniquely determine
the parameters $C$,  $D$ and $K$.  However, in  principle, with a good
enough  data  set, one  could  determine  $K$  uniquely from  equation
(\ref{K}). Then, since $K \gg C$, one could use equation (\ref{total})
to determine the parameter $D$. This will then give a determination of
the redshift evolution of the  IBL independently of models of the star
formation rate.

\begin{deluxetable}{cccc}
\tabletypesize{\small}
%%\rotate
\tablecaption{Steepening   Parameters}   \tablewidth{0pt}  \tablehead{
\colhead{Evolution Model} & \colhead{C}  & \colhead{D} & \colhead{K} }
\startdata  Fast Evolution  &  -0.0972  & 10.6  &  0.427\\ Baseline  &
-0.0675 & 7.99 & 0.716 \\ \enddata
\end{deluxetable}

\begin{figure}[h]
\epsscale{.80} \plotone{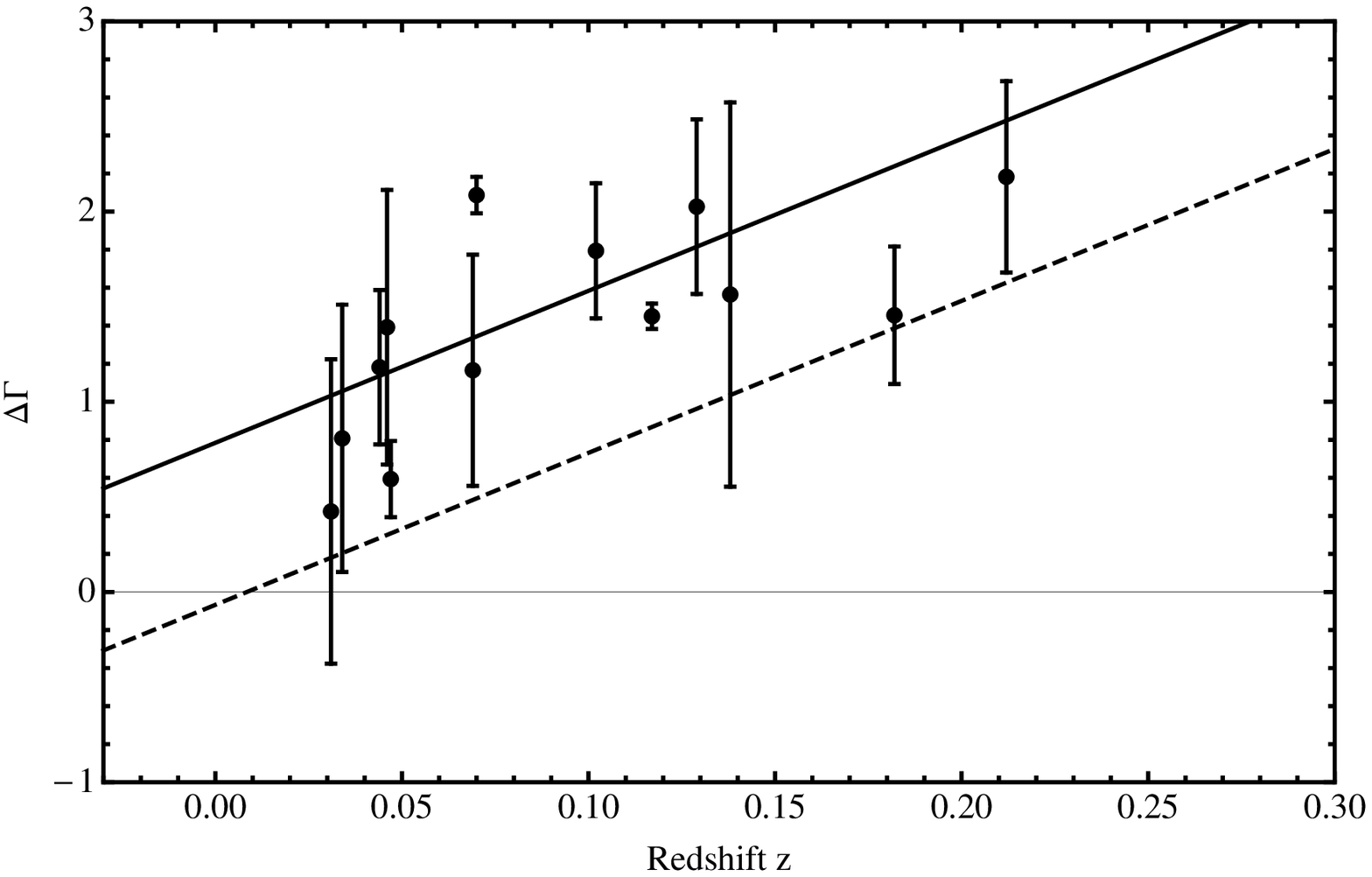}
\caption{The fits obtained  for the linear functions $C  + Dz$ (dashed
line) and $C + Dz + K$ (solid line) shown for the SMS06 baseline model
as descibed  in the text.   These are  fit to the  data on 13  BL Lacs
given by Abdo et al. (2009).
\label{f1}}

\end{figure}

\begin{figure}[h]
\epsscale{.80} \plotone{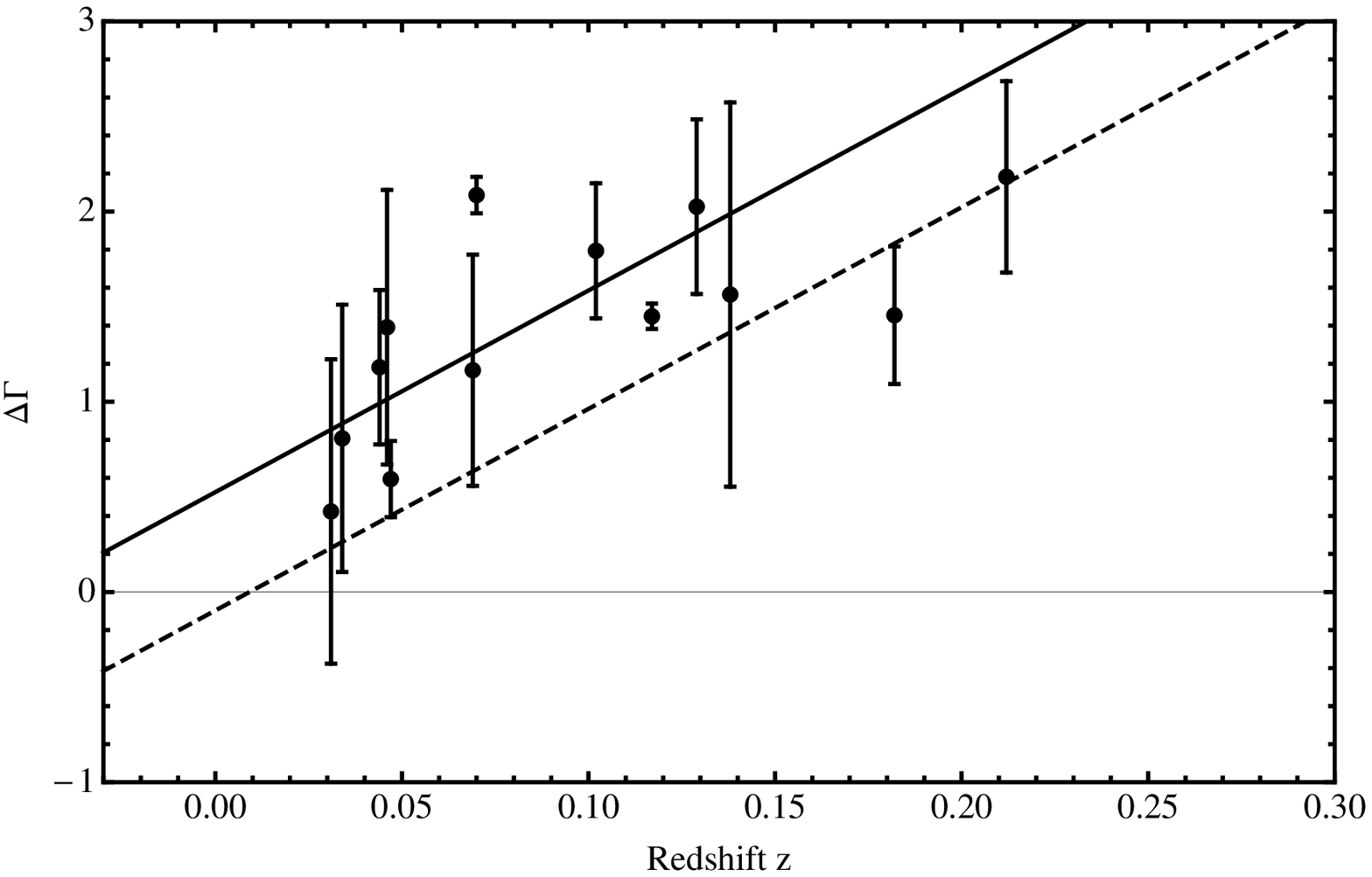}
\caption{ The fits obtained for  the linear functions $C + Dz$ (dashed
line) and $C + Dz + K$ (solid line) shown for the SMS06 fast evolution
model as  descibed in the text.   These are fit  to the data on  13 BL
Lacs given by Abdo et al. (2009).
\label{f2}}

\end{figure}

Figures \ref{f1} and \ref{f2} show the fits of the parameters given in
Table 1  to the  linear dependence in  and redshift given  by equation
(\ref{total}).

\section{Conclusions}

We have  derived a simple  analytic approximation for  determining the
steepening in the spectra of TeV  selected BL Lac AGN and compared our
results with recent observational data  on 13 TeV selected BL Lac AGN.
Our relation is in excellent agreement with the observational data and
provides a framework for understanding and interpreting both these and
observational data.

SS06 have  shown that  the effect of  intergalactic absorption  on the
spectra of AGN in the energy range 0.2 TeV $ < E_{\gamma} <$ 2 TeV and
the redshift range $0.05 < z < 0.4$ is a simple power-law to power-law
steepening.  Absorption in this  energy range is primarily produced by
interactions with near  infrared photons from on the  low energy sides
of  the  starlight peaks  in  galaxy SEDs.   The  {\it  shape} of  the
resulting  peak in  the  intergalactic SED  produces an  approximately
logarithmic  energy dependence for  the function  $\tau (E_{\gamma})$.
This energy  dependence should be roughly  the same for  all models of
the IBL.   Therefore, our general  result of a linear  $\Delta\Gamma =
(C+K) + Dz$ relationship is  roughly independent of the specific IBL
model used  because  it only
depends on the shape of the average  galaxy SED on the near IR side of
the IBL starlight  peak. The parameters $C,  D$ and $K$  will be different
for the  different models  since the absolute  value of  $\tau$ varies
from one IBL model to another.

Given our derived relation between $\Delta \Gamma$ and redshift, 
and with good enough data on the mean \gray SED of TeV Selected BL Lacs
used to determine $K$, the redshift evolution of the IBL can, in principle, 
be determined independently of stellar evolution models.

\section*{Acknowledgment} We thank Stephen Fegan and David Sanchez 
for sending the BL Lac data table used in our Figures 1 and 2. We thank
Markos Georganopoulos for a helpful discussion.

\end{document}